\documentclass[aps,twocolumn]{revtex4}
\usepackage{psfig}
\usepackage{graphics}
\usepackage{amsfonts,amssymb}
\begin{document}

\title{Classical Theory of Optical Nonlinearity in Conducting
  Nanoparticles}

\author{George Y. Panasyuk, John C. Schotland} 

\affiliation{Department of Bioengineering, University of Pennsylvania,
  Philadelphia, PA 19104}

\author{Vadim A. Markel} 

\affiliation{Departments of Radiology and Bioengineering, University of Pennsylvania, Philadelphia, PA 19104}

\begin{abstract}
  We develop a classical theory of electron confinement in conducting
  nanoparticles. The theory is used to compute the nonlinear optical
  response of the nanoparticle to a harmonic external field.
\end{abstract}

\date{\today} 

\maketitle 

Fundamental and applied research in the area of plasmonic nanodevices
and of engineered materials constructed from plasmonic nanoparticles
is at the center of modern
optics~\cite{shalaev_book_07,brongersma_book_07}. Most theoretical
approaches to such systems are based on the classical electrodynamics
of continuous media. However, in the case of nanoparticles whose
linear dimensions are not dramatically larger than the atomic scale (a
factor of $\sim 10$ is typical), finite-size and quantum corrections
are of importance~\cite{kreibig_85_1}.  Previously, the above two
terms have been used interchangeably and it is accepted that the
small-size effects are quantum mechanical in origin. Thus, they can be
understood by considering discrete electron states in a
nanoparticle~\cite{hache_86_1,rautian_97_1} or reduction of interband
screening and electron spill-over near the nanoparticle
surface~\cite{pustovit_06_1,pustovit_06_2}. In this Letter, we
demonstrate that there is an additional, purely classical mechanism
that leads to finite-size effects and, in particular, to nonlinearity
of the electromagnetic response of conducting nanoparticles. We refer
to this mechanisms as the {\em classical confinement effect}.

In this work, we develop a theory of classical confinement of
electrons in conducting subwavelength-sized nanoparticles and derive
nonlinear polarizabilities in 1D (slab) and 3D (sphere) geometries.
Our theory is nonperturbative and fully accounts for electron-electron
interactions within the accuracy of the quasistatic approximation. The
predicted effects differ from other optical nonlinearities, most
importantly, by the unusual dependence of the nonlinear response on
the intensity of the incident laser beam. 

Optical nonlinearities in noble metal nanoparticles in different
phases (such as dilute solutions, random composites and fractal
aggregates) have been extensively investigated (see
Refs.~\cite{drachev_04_2,torres-torres_07_1} and references therein).
Most experimental studies were focused on the influence of the
composite geometry on the nonlinear optical
response~\cite{butenko_90_1,danilova_96_2,danilova_97_1,drachev_98_1,karpov_01_1}.
Measurements of the nonlinear response as a function of the intensity
of the incident beam have been largely limited to nonlinear refraction
and absorption. We note that the experimental data of
Ref.~\cite{danilova_97_1} for the dependence of the nonlinear
correction to the absorption coefficient on the incident beam
intensity are poorly explained and more in line with the theory
developed below.

On the theoretical side, the size-dependent nonlinear susceptibilities
of a conducting nanoparticle were calculated in
Refs.~\cite{hache_86_1,rautian_97_1} from the first principles. A
conducting nanosphere was modeled as a degenerate electron gas placed
in an infinitely-high confining potential and subjected to a
time-harmonic external electric field. In this model, the optical
nonlinearity is a consequence of the saturation of optical transitions
between discrete electronic states, however, the confinement effect
which we discuss in this Letter is not taken into account. Thus, for
example, in Ref.~\cite{rautian_97_1}, a Hamiltonian was used with the
interaction term $V = e{\bf r}\cdot {\bf E}$ where ${\bf E} = f{\bf
  E}_{\rm ext}$, ${\bf E}_{\rm ext}$ being the external field and $f =
[1 + (\epsilon - 1)/3]^{-1}$ the Lorentz factor. In other words, it
was assumed that the electrons move in an electric field that is
obtained from the solution to the Laplace equation {\em which does not
  account for finite-size effects}. In a more recent theoretical
work~\cite{pustovit_06_1,pustovit_06_2}, a finite-size effect which is
conceptually related to the one described in this Letter but occurs in
a very different physical setting was considered for the purpose of
calculating the enhancement factor of Raman scattering by a molecule
adsorbed on the surface a nanoparticle.  However, polarization of the
nanoparticle itself was not addressed
in~\cite{pustovit_06_1,pustovit_06_2}. In contrast to
Refs.~\cite{hache_86_1,rautian_97_1,pustovit_06_1,pustovit_06_2}, we
present here an analytical and fully self-consistent theory of the
nonlinear optical response of conducting nanoparticles.

We begin by noting that the classical electrodynamic theory of
conductors is based on the implicit assumption that the volume density
of free charge is infinite. More specifically, the atomic lattice is
assumed to be rigid and to carry a uniform positive volume charge
while free electrons form a negatively-charged compressible fluid. If
we apply an external field $E_{\rm ext}$, the two volume charges shift
with respect to each other by a distance $\delta$ which results in the
formation of a surface charge with density $\sigma$. From the
linearity of Laplace equation, it follows immediately that $\sigma
\propto \delta\propto E_{\rm ext}$. However, the volume charge
densities are assumed to be so large that, irrespective of the
magnitude of the external field, the shift $\delta$ is much smaller
than any other physical scale in the problem. This assumption is
exceedingly accurate for macroscopic conductors. But in nanoparticles,
a nonzero value of $\delta$ can result in experimentally observable
nonlinearity.

As a simple qualitative illustration, consider a conducting sphere of
radius $a$ in an external time-varying field $\tilde{E}_{\rm ext} =
E_0\exp(-i\omega t)$ polarized along the $z$-axis. Here and
thereafter, we denote the complex representation of physical
observables by a tilde; the corresponding real quantities are obtained
by adding a complex conjugate, i.e., $E = \tilde{E} + \tilde{E}^*$.
Within quasistatics, the conventional result is that the sphere
acquires a surface charge density $\tilde{\sigma} = (3E_0 /4\pi)
[(\epsilon-1)/(\epsilon+2)] \cos\theta \exp(-i\omega t)$, where
$\theta$ is the polar angle and $\epsilon=\epsilon(\omega)$ is the the
permittivity of the conductor. At low frequencies, $\epsilon
\rightarrow i\infty$ and $\tilde{\sigma} = (3E_0/4\pi)\cos\theta
\exp(-i\omega t)$. We can use this expression to compute the dipole
moment $\tilde{d}$ of the sphere, to obtain $\tilde{d} = a^3 E_0
\exp(-i\omega t)$.  Now let $E_0>0$ and consider such values of the
time $t$ and the angle $\theta$ that $\cos\theta \exp(-i\omega t) =
1$.  Then the real-valued surface charge density is $\sigma =
2\tilde{\sigma} = 3E_0 / 2\pi > 0$. Since the lattice is viewed as
rigid, the above surface charge is produced by a layer of depth $h =
3E_0/2\pi \rho$, where $\rho = Ze/\ell^3$ is the volume charge density
of the lattice, $Z$ is the number of free electrons per atom, $e$ is
the electron charge, $\ell$ is the lattice constant, we have assumed a
cubic lattice and neglected surface roughness.  We thus find that $h/a
= (3/4\pi)(E_0 / E_{\rm at}) (\ell / a)$, where $E_{\rm at} =
Ze/\ell^2$ is the atomic field. Once we take into account the nonzero
value of $h$, the linear formula $\tilde{d} = a^3 E_0 \exp(-i\omega
t)$ is no longer valid. We emphasize that the above considerations
applies only to the positive charge. In the classical confinement
model, a layer of negative surface charge with negligibly small depth
can still be formed since the conduction electrons are viewed as a
compressible fluid.

Now consider the magnitude of the effect described above. The theory
has two small parameters: $E_0/ E_{\rm at}$ and $\ell / a$.  The first
parameter is typical in nonlinear optics~\cite{boyd_book_92}. The
second small parameter, $\ell/a$, is negligibly small for macroscopic
spheres.  However, for $a \sim 5\ {\rm nm}$ and $\ell \sim 0.5\ {\rm
  nm}$ (silver), the ratio is $\sim 1/10$. Further, we will see below
that the nonlinearity in the classical confinement model is manifested
to the second order in $E_0$, while in the quantum-mechanical theory
of Refs.~\cite{hache_86_1,rautian_97_1}, the first nonlinear
correction is in the third order. We thus conclude that the
nonlinearity described in this Letter is of the same order or larger
than other typical optical nonlinearities.

It can be argued on very general grounds that the mathematical
properties of the nonlinear polarizability in the classical
confinement model are markedly different from that in the
quantum-mechanical model where nonlinearity is a consequence of the
saturation of electronic transitions. To understand this, consider
the polarization of a two-level quantum system with energy gap $\hbar
\omega_{21}$ and relaxation constants $\Gamma_{mn}$ which is pumped
with the rates $Q_m$ ($m,n=1,2$; the terms $\Gamma_{mn}$ describe
relaxation rates for the corresponding elements of the density matrix
$\varrho$; in a closed system $Q_1 = \Gamma_{22}\varrho_{22}$) and is
excited by a time-harmonic field $\tilde{\bf E}_{\rm ext} = {\bf E}_0
\exp(-i\omega t)$.  The expectation value of the dipole moment is
$\langle \tilde{\bf d} (t) \rangle = {\bf d}_{12} {\mathcal R}_{21}
\exp(-i \omega t)$, where ${\bf d}_{12}$ is the dipole moment matrix
element and ${\mathcal R}_{21}$ is the slowly-varying amplitude of the
off-diagonal element of the density matrix. In the rotating-wave
approximation, the latter is given by

\begin{equation}
\label{alpha_QM}
{\mathcal R}_{21} = \frac{G}{\Omega - i \Gamma_{21}}
\frac{Q_2/\Gamma_{22} - Q_1/\Gamma_{11}}{1 + 2\kappa (1/\Gamma_{22} +
  1/\Gamma_{11})} \ ,
\end{equation}

\begin{equation}
\Omega = \omega_{21} - \omega \ , \ \ 
G = - {\bf d}_{21} \cdot {\bf E}_0 \ , \ \ 
\kappa = \frac{\vert G \vert^2 \Gamma_{21}}{\Omega^2  +  \Gamma_{21}^2}
\ .
\end{equation}

\noindent
If the incident wave is linearly polarized~\cite{fn_2}, we have $\vert
G \vert^2 = \vert {\bf d}_{21} \vert^2 E_0^2$. Thus, the expectation
of the dipole moment, as well as the polarizability tensor, are
meromorphic functions of the field amplitude $E_0$. In particular,
$\langle \tilde{\bf d}(t) \rangle$ can be expanded into a Taylor
series in $E_0$. Due to the spherical symmetry of the problem, this
expansion contains only odd powers of the field.  As is the case for
any meromorphic function whose Taylor expansion does not contain a
zeroth-order term, the dipole moment vanishes in the limit $E_0
\rightarrow \infty$.  Physically, this corresponds to saturation of
the transition. Indeed, the second factor in the right-hand side of
Eq.~(\ref{alpha_QM}) describes field-dependent population inversion
which, in the limit of a strong incident field, approaches zero.  The
saturation process is much more complicated in the case of a
degenerate electron gas in a spherical potential well which was
considered in Refs.~\cite{hache_86_1,rautian_97_1}, yet the analytical
properties of the solutions are essentially the same.

In the case of classical confinement, the situation is substantially
different.  Here the dipole moment does not vanish in the limit of
strong external fields but rather oscillates with an amplitude that
approaches a constant limit. Consider again a conducting sphere in a
time-harmonic external field. We will ignore the possibility of
ionization, i.e., similarly to Refs.~\cite{hache_86_1,rautian_97_1},
we will assume the existence of an infinite potential barrier that
prevents electrons from leaving the system. Then, for a very strong
external field, the amplitude of the oscillating dipole moment is
$d_{\rm max} = Nea$, $N$ being the number of free electrons in the
system. Here $d_{\rm max}$ is the dipole moment which is obtained when
all conducting electrons are concentrated at a point near the sphere's
surface. Although the limit we have just considered is arguably
unphysical, its mathematical implications are important. Indeed, there
is no such analytic function of $E_0$ that is zero at $E_0=0$ and
approaches a finite limit at infinity. We conclude that the function
$d(E_0)$ is not analytic and can contain, for example, a term $\propto
E_0 \vert E_0 \vert$ which, although it is second order in $E_0$, does
not need to vanish in centrosymmetric systems. We will see below that
this is, indeed, the case.

We now proceed with more detailed calculations. As a first step,
consider the one-dimensional problem schematically illustrated in
Fig.~1a. Here an external field $E_{\rm ext} = - E_0$ is directed
perpendicularly to a slab of thickness $L$. If $E_0>0$, the free
charge is distributed inside the slab as follows. The surface which is
opposite the field direction (as shown in the figure) acquires a
negative surface charge $\sigma_2 = - E_0/4\pi$. Near the other
surface, a positively-charged layer of depth $h = E_0 / 4\pi\rho$ is
formed, where $\rho = Ze/\ell^3$ is the background positive volume
charge density. We thus see that the slab is separated into two
regions. The first region is characterized by zero conductivity due to
the absence of free carriers (this region is dashed in Fig.~1a) while
the second region is conducting. In static equilibrium, the local
field in the second region must be zero, while there is no such
requirement for the first region. We note that the field inside the
conducting region which is produced by the positively charged layer is
$2\pi\sigma_1$ where $\sigma_1 = E_0 /4\pi$, the same as would be
produced by a surface charge of density $\sigma_1$. Thus, the
depolarizing field inside the conducting layer is $E_{\rm dep} =
2\pi(\sigma_1 - \sigma_2) = E_0$ and the total local field $E_{\rm
  loc} = E_{\rm ext} + E_{\rm dep}$ is zero.\\

%\begin{figure}
\centerline{\psfig{file=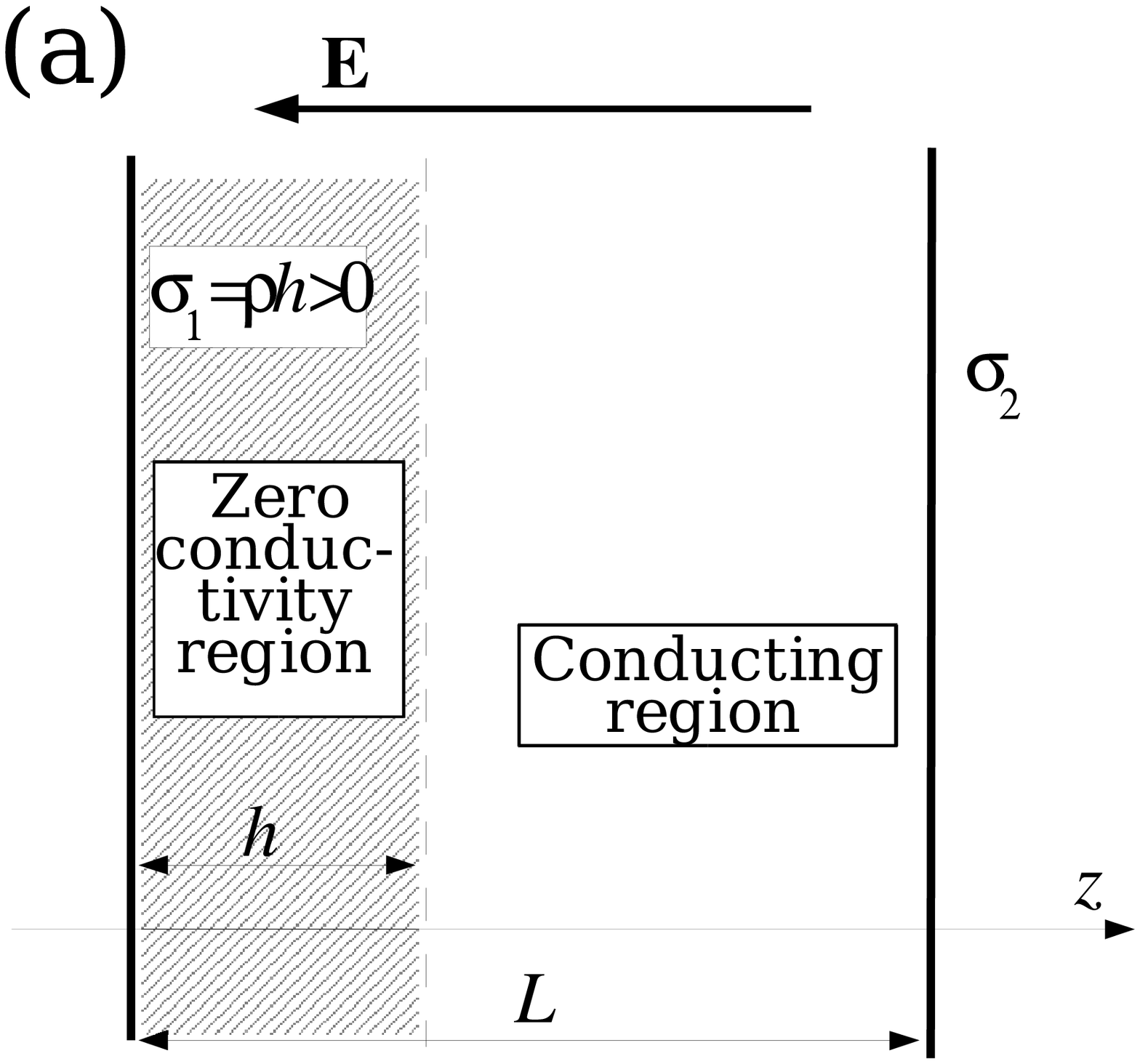,width=6.cm,bbllx=50bp,bblly=200bp,bburx=550bp,bbury=760bp,clip=t}}
%\vspace*{-3mm}
\centerline{\psfig{file=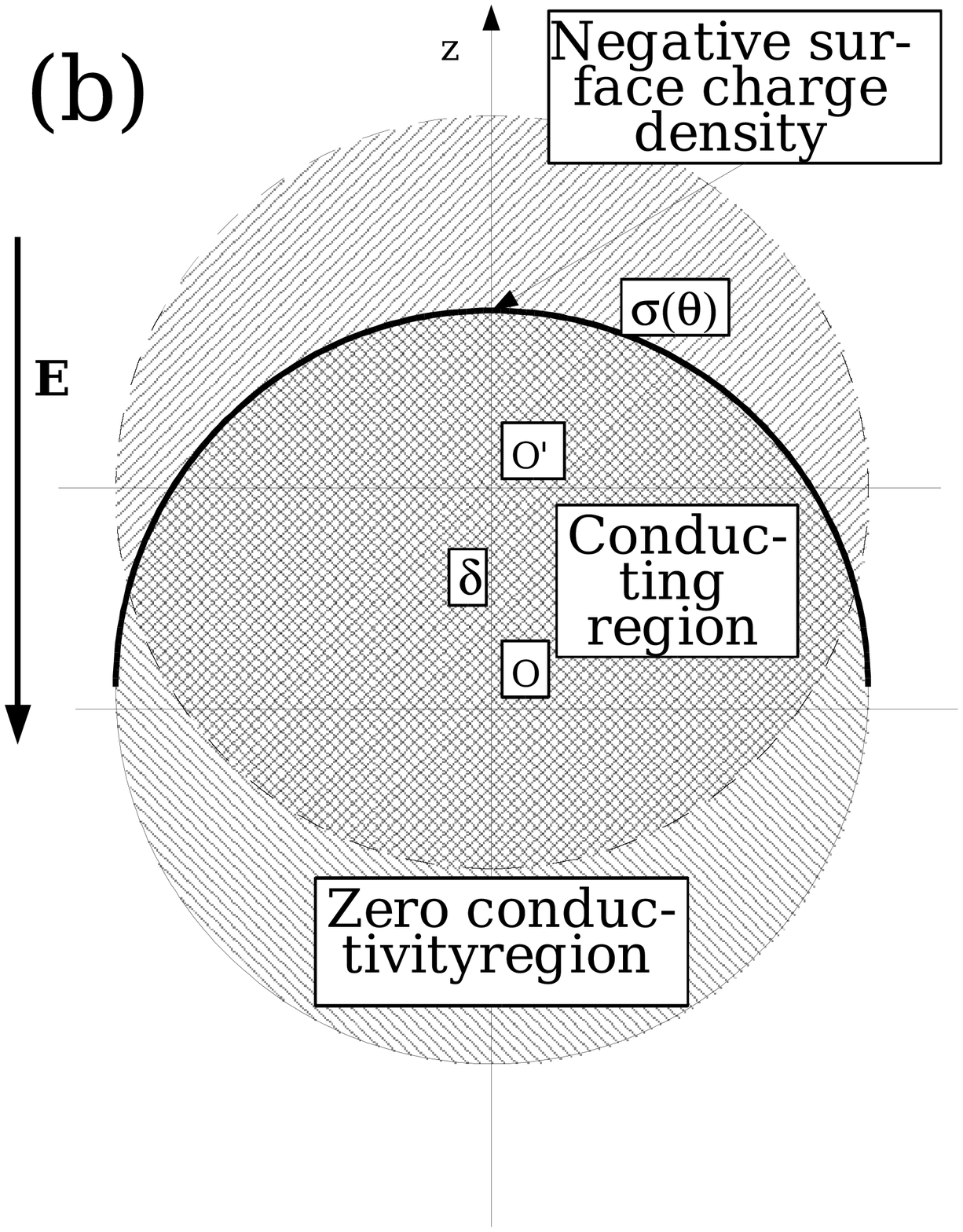,width=6.cm,bbllx=50bp,bblly=100bp,bburx=550bp,bbury=700bp,clip=t}}
{\small FIG.~1. Illustrating the geometry of the 1D (a) and 3D (b) problems.}\\
%\end{figure}

Next, consider a time-varying external field $\tilde{E}_{\rm ext}(t) =
E_0 \exp(-i\omega t)$. The positively-charged layer can now appear on
either side of the slab. It is convenient to introduce the following
notation: let the total charge per unit area which accumulates near
the left face of the slab be $\sigma_1$ and the similar quantity near
the right face be $\sigma_2$. Here we do not distinguish between a
true surface charge and a surface layer of finite depth. It follows
from charge conservation that $\sigma_1 = - \sigma_2$.  We then can
write $\sigma_1 = -\sigma$, $\sigma_2 = +\sigma$, where $\sigma$ can
be either positive or negative, depending on the phase of the
oscillations.  The depolarizing field in the conducting region is
given at any time by $E_{\rm dep}=-4\pi\sigma$ and the total local
field by $E_{\rm loc} = E_{\rm ext} + E_{\rm dep}$.  Then the equations
of motion for a negative test charge inside the conducting region can
be written as

\begin{equation}
\label{motion_1D}
m(\dot{v} + \gamma v) = -e[E_{\rm ext}(t) - 4\pi\sigma ] \ , \ \ \dot{\sigma} = -\rho v \ .
\end{equation}

\noindent
Here $v, m$ and $e$ are the electron velocity, mass and charge,
respectively, and $\gamma$ is a phenomenological friction term; the
first equation in (\ref{motion_1D}) is Newton's second law while the
second is the continuity equation. The oscillatory solution to the
above system of ordinary differential equations is

\begin{equation}
\label{sigma_1D_solution}
4\pi \tilde{\sigma}(t) = \omega_p^2 E_0 e^{-i\omega t} \left[\omega_p^2 - \omega^2 -
  i\gamma\omega \right]^{-1} \ ,
\end{equation}

\noindent
where $\omega_p=\sqrt{4\pi e\rho / m}$ is the plasma frequency.  We
can further compute the local field $E_{\rm loc}$ and the current $j =
-\rho v$ inside the conducting region and verify that the ratio
$\tilde{j}(t)/\tilde{E}_{\rm loc}(t)$ yields the classical Drude
conductivity $i \omega_p^2/4\pi(\omega + i\gamma)$.
 
So far, the results appear to be conventional. Finite-size effects and
the nonlinearity of the optical response become apparent when we
compute the dipole moment per unit area of the slab, ${\mathcal
  P}=\Delta d / \Delta S$. A straightforward calculation yields
${\mathcal P}(t) = \sigma(t)[L - h/2]$. Thus, the effective width of
the slab is reduced by $h/2$ where $h = \vert \sigma \vert /\rho$ (note
that $h$ is related to the absolute value of the {\em real-valued}
quantity $\sigma$). We now find that ${\mathcal P}(t) = \sigma(t)[L -
(1/2\rho) \vert \sigma(t) \vert]$. We further note that the field
amplitude $E_0$ can always be chosen to be real so that $\sigma(t) =
E_0 A(\omega) \cos(\omega t - \varphi)$, where $A(\omega) =
\omega_p^2/2\pi\sqrt{(\omega_p^2 - \omega^2)^2 + (\gamma\omega)^2 }$
and $\tan \varphi = \gamma\omega/(\omega_p^2 - \omega^2)$. We then
write the final result as

\begin{equation}
\label{P_1D}
{\mathcal P}(t) = \alpha E_0 \cos(\omega t - \varphi) \Big [ 1 + \beta
\big \vert E_0 \cos(\omega t - \varphi) \big \vert \Big ] \ ,
\end{equation}

\noindent
where $\alpha = L A(\omega)$, $\beta = - A(\omega)/2\rho L =
-A(\omega)(\ell/2L) E_{\rm at}^{-1}$ and we have introduced again the
microscopic quantities $\ell=(Ze/\rho)^{1/3}$ and $E_{\rm at} =
Ze/\ell^2$. 

Note that (\ref{P_1D}) is not an expansion but is exact as long as $h
< L$ or, equivalently, $\max(\vert \sigma \vert) < \rho L$. The
important feature of the obtained solution is that ${\mathcal P}(E_0)$
is not an analytic function and can not be expanded into a Taylor
series. This mathematical property is closely related to the existence
of a finite limit $\lim_{E_0 \rightarrow \infty} {\mathcal P}(E_0)$
(in the saturation model of Refs.~\cite{hache_86_1,rautian_97_1} this
limit is zero). The non-analyticity of the optical response has
important consequences for the generation of frequency harmonics as is
shown below for the more physically important 3D case.

We now consider the problem of a three-dimensional conducting sphere.
It turns out that accounting for classical confinement in this case
leads to a formidable mathematical problem. We will use, however,
certain physical insights that will allow us to obtain a
nonperturbative analytical theory. Consider a conducting sphere of
radius $a$, a constant background positive volume charge $\rho$ and a
free charge whose integral over the sphere volume is $-4\pi a^3 \rho
/3$; an infinitely high spherical potential barrier that prevents
ionization is assumed. We seek to find the dipole moment of the sphere
in an external field $\tilde{\bf E}_{\rm ext} = \hat{\bf
  z}E_0\exp(-i\omega t)$, where $\hat{\bf z}$ is a unit vector
pointing in the direction of the $z$-axis. As before, we recognize
that the sphere is separated at all times into two regions: one region
has no free carriers and is nonconducting while the other region has a
constant nonzero conductivity; this conducting region is doubly dashed
in Fig.~1b.

The first physical observation that we make is that in the quasistatic
problem with a time-harmonic external field, the motion of charges is
such that, at any time $t$, both the volume and the surface charge
densities correspond to a static equilibrium which is obtained for an
external field $A(\omega) {\bf E}_{\rm ext}(t^{\prime})$ which is
taken {\em at a different time} $t^{\prime}$ and multiplied by a
frequency-dependent real-valued factor $A(\omega)$. Thus, the system
goes through states of static equilibrium which are phase-shifted with
respect to the external field. Mathematically, this statement follows
from the linearity of the equations of motion. In static equilibrium,
the electric field in the conducting region is zero. In the dynamic
problem, the latter is nonzero but proportional to the difference
${\bf E}_{\rm ext}(t) - {\bf E}_{\rm ext}(t^{\prime})$. We assume here
that ${\bf E}_{\rm ext}$ is spatially homogeneous over the volume of
the sphere. The only motion of the free charge {\em inside the
  conducting region} that is consistent with this condition is
one-dimensional motion along the $z$-axis. From this, we find that the
surface that separates the conducting and non-conducting regions is a
sphere. The center of this sphere is denoted by $O^{\prime}$ and is
shifted from the center of the original sphere by a distance $\delta$
along the $z$-axis, where $\delta$ can be both positive and negative
(see illustration in Fig.~1b). We thus can characterize the dynamics
of the system by a single scalar parameter $\delta$.

The second observation will allow us to find the depolarizing field
inside the conducting region. As we have argued above, the total
electric field inside that region is spatially homogeneous and
directed along the $z$-axis. The external field does satisfy this
condition and so must the depolarizing field ${\bf E}_{\rm dep}$. The
latter is a superposition of the field produced by a positively
charged meniscus and the negative surface charge $\sigma$ which we
have not yet determined. We notice, however, that a field with the
required properties is created by two oppositely charged menisci of
the same shape which are shown in Fig.~1b as single-dashed regions.
Indeed, the field produced by the two menisci is the same as the field
of two oppositely charged spheres shifted with respect to each other
by a distance $\delta$. Inside the conducting region, this field is
given by ${\bf E}_{\rm dep} = 4\pi\rho\delta \hat{\bf z} / 3$. With
the understanding that this field is created by the positively charged
meniscus and by a yet unknown negative surface charge $\sigma$ {\em
  whose field in the conducting region is the same as that of the
  hypothetical negatively-charged meniscus}, we write the equation of
motion as

\begin{equation}
\label{motion_3D}
m\left(\ddot{\delta} + \gamma \dot{\delta}\right) = -e \left [\hat{\bf
    z}\cdot {\bf E}_{\rm ext}(t) + 4\pi\rho\delta/3 \right] \ .
\end{equation}

\noindent
The oscillatory solution to (\ref{motion_3D}) is

\begin{equation}
\label{delta_solution}
\tilde{\delta}(t) = -(e/m)E_0e^{-i\omega t} \left[\omega_F^2 -
  \omega^2 - i\gamma\omega \right]^{-1} \ ,
\end{equation}

\noindent
where $\omega_F = \omega_p/\sqrt{3}$ is the Frohlich frequency. In the
conventional approach, the dipole moment of the sphere is computed as
$\tilde{d} = - 4\pi\rho a^3 \tilde{\delta}/3$. Evaluation of this
expression leads to the linear polarizability $\alpha =
a^3(\epsilon-1)/(\epsilon+2)$ with $\epsilon = 1 -
\omega_p^2/\omega(\omega+i\gamma)$. We, however, intend to take into
account the presence of the meniscus and the fact that the surface
charge density can deviate from the usual $\propto\cos\theta$
dependence. To this end, we write $d_z = \hat{\bf z} \cdot{\bf d} =
\rho \int_V z d^3r + \int_S z \sigma d^2 r$. Here the first integral
is over the volume of the positively charged meniscus while the second
term is the contribution of the negative surface charge. After tedious
but straightforward integration, we obtain

\begin{equation}
\label{d_z}
\frac{d_z}{2\pi} = -\frac{\rho \delta}{3}\left[a^3 - \vert \delta \vert
  \left(\frac{3a^2}{4} - \frac{\delta^2}{16}\right) \right] + a^3 \sigma_1 \ ,
\end{equation}

\noindent
where $\sigma_1 = \int \sigma(\cos\theta) \cos\theta d\cos\theta$. We
thus see that knowledge of the first moment of the surface charge
density suffices for the purpose of computing the dipole moment. While
finding the whole function $\sigma(\cos\theta)$ is a difficult task,
$\sigma_1$ can be found by considering the depolarizing potential in
the vicinity of the sphere origin, $O$. On one hand, we know that the
potential is given by the formula $\phi = -E_{\rm dep} z$ with $E_{\rm
  dep} = 4\pi\rho\delta/3$. On the other hand, we can write the same
potential as an integral over the meniscus and the surface of the
sphere and expand the resultant formula into scalar spherical
harmonics $r^l Y_{lm}(\hat{\bf r}\cdot \hat{\bf z})$. A
straightforward calculation of the term $l=1,m=0$ yields $\phi_{10} =
2\pi[\sigma_1 - (\rho \delta/3)(1 + (3\vert \delta \vert/8 a))] z$.
[The condition that all higher terms in the above expansion vanish
provides an infinite set of equations for the higher moments of
$\sigma(\cos \theta)$.] From the equality $\phi_{\rm dep} =
\phi_{10}$, it follows that

\begin{equation}
\label{sigma_1}
\sigma_1 = - (\rho\delta/3)\left[1 - (3 \vert \delta \vert / 8a) \right ] \ .
\end{equation}

\noindent
We then substitute the above result into (\ref{d_z}) to find

\begin{equation}
\label{d_z_delta}
d_z = -\frac{4\pi\rho\delta a^3}{3}\left[1 - \frac{9}{16}\frac{\vert
    \delta \vert}{a} + \frac{1}{32}\frac{\delta^2\vert \delta
    \vert}{a^3} \right]
\end{equation}

\noindent
where $\delta = \tilde{\delta} + \tilde{\delta}^*$ and
$\tilde{\delta}$ is given by (\ref{delta_solution}). We emphasize that
the above formula is not an expansion: it is exact as long as $\vert
\delta \vert < a$, the latter condition having been implicitly used
for computing $\sigma_1$.

We thus find that the polarizability of the sphere has a mathematical
structure which is similar to that of the one-dimensional slab, except
for the additional term $\propto \delta^3\vert \delta \vert$ which
appears in (\ref{d_z_delta}). For most practical purposes, this term
can be neglected since it contains an additional small factor $(E_0
\ell/E_{\rm at}a)^2$ compared to the term $\propto \delta
\vert\delta\vert$.  Both terms, however, describe the generation of
odd-order frequency harmonics, i.e., oscillations at the frequencies
$\omega_n = (2n+1) \omega$, $n=0,1,2,\ldots$. Indeed, let $E_0$ be
purely real, so that $\delta \vert \delta \vert \propto \cos(\tau)
\vert \cos(\tau)\vert$ and $\delta^3 \vert \delta \vert \propto
\cos^3(\tau) \vert \cos(\tau)\vert$, where $\tau = \omega t -
\varphi$, and consider the following Fourier series:

\begin{eqnarray}
\label{delta_2}
\cos(\tau) \big\vert
\cos (\tau) \big\vert = \sum_{k=1}^{\infty} 
\frac{8\sin(\pi k/2)}{\pi k(4-k^2)} \cos(k\tau)
\\
\label{delta_4}
\cos^3(\tau) \big\vert
\cos (\tau) \big\vert = \sum_{k=1}^{\infty} 
\frac{96\sin(\pi k/2)}{\pi k(64 - 2k^2 + k^4)} \cos(k\tau)
\end{eqnarray}

\noindent
In both cases the Fourier coefficients are nonzero only if $k=2n+1$
(the $k=2$ Fourier coefficient in Eq.~(\ref{delta_2}) is zero despite
the existing uncertainty). Note that even though the first nonlinear
correction is second order in $E_0$, there is no second-harmonic
generation, which is also the case in the traditional theory of
nonlinear polarization of centrosymmetric systems. However, generation
of the third harmonic ($n=1$) and nonlinear refraction ($n=0$) which
are traditionally associated with the third-order nonlinear
susceptibility $\chi^{(3)}$, are manifest in the model to the second
order in $E_0$. Note that higher-order harmonics may have vacuum
wavelengths which are short enough to render the quasistatic
approximation invalid.

In summary, we have described novel nonlinear optical effects in
conducting nanoparticles and developed a quantitative theory of
nonlinear polarization of a conducting nanosphere
(Eqs.~(\ref{delta_solution}),(\ref{d_z_delta})-(\ref{delta_4})).
Experimentally, the predicted effects can be distinguished from other
optical nonlinearities by investigating the dependence of the
nonlinear response on the intensity $I_0$ of the incident laser beam.
Thus, the nonlinear correction to the refractive index is predicted to
scale as $\propto \sqrt{I_0}$. The intensity of frequency harmonics
generated due to the classical confinement effect scales as $\propto
I_0^2$, irrespective of the harmonic order.

This research was supported by the NSF under Grant DMR 0425780 and by
the AFOSR under Grant FA9550-07-1-0096. The contact e-mail is
vmarkel@mail.med.upenn.edu.  

\bibliographystyle{prsty}
\vspace{-2mm}
\bibliography{abbrev,master,book,local}

\begin{thebibliography}{16}
\expandafter\ifx\csname natexlab\endcsname\relax\def\natexlab#1{#1}\fi
\expandafter\ifx\csname bibnamefont\endcsname\relax
  \def\bibnamefont#1{#1}\fi
\expandafter\ifx\csname bibfnamefont\endcsname\relax
  \def\bibfnamefont#1{#1}\fi
\expandafter\ifx\csname url\endcsname\relax
  \def\url#1{\texttt{#1}}\fi
\expandafter\ifx\csname urlprefix\endcsname\relax\def\urlprefix{URL }\fi
\providecommand*{\bibinfo}[2]{#2}
\providecommand*{\eprint}[1]{#1}
\providecommand*{\url}[1]{#1}
\begingroup\makeatletter
 \@temptokena{%
  \expandafter\ifx\csname citenamefont\endcsname\relax
   \DeclareRobustCommand\citenamefont{\@firstofone}%
   \global\let\citenamefont\citenamefont
   \global\expandafter\let\csname citenamefont \expandafter\endcsname\csname
  citenamefont \endcsname
  \fi
 }\if@filesw\immediate\write\@auxout{\the\@temptokena}\fi
\expandafter\endgroup\the\@temptokena

\bibitem[{\citenamefont{Shalaev and Kawata}(2007)}]{shalaev_book_07}
\bibinfo{editor}{\bibfnamefont{V.~M.} \bibnamefont{Shalaev}} \bibnamefont{and}
  \bibinfo{editor}{\bibfnamefont{S.}~\bibnamefont{Kawata}}, eds.,
  \emph{\bibinfo{title}{Nanophotonics with Surface Plasmons}}
  (\bibinfo{publisher}{Elsevier}, \bibinfo{address}{Amsterdam},
  \bibinfo{year}{2007}).

\bibitem[{\citenamefont{Brongersma and Kik}(2007)}]{brongersma_book_07}
\bibinfo{editor}{\bibfnamefont{M.~L.} \bibnamefont{Brongersma}}
  \bibnamefont{and} \bibinfo{editor}{\bibfnamefont{P.~G.} \bibnamefont{Kik}},
  eds., \emph{\bibinfo{title}{Surface Plasmon Nanophotonics}}
  (\bibinfo{publisher}{Springer}, \bibinfo{address}{Dordrecht},
  \bibinfo{year}{2007}).

\bibitem[{\citenamefont{Kreibig and Genzel}(1985)}]{kreibig_85_1}
\bibinfo{author}{\bibfnamefont{U.}~\bibnamefont{Kreibig}} \bibnamefont{and}
  \bibinfo{author}{\bibfnamefont{L.}~\bibnamefont{Genzel}},
  \bibinfo{journal}{Surface Science} \textbf{\bibinfo{volume}{156}},
  \bibinfo{pages}{678} (\bibinfo{year}{1985}).

\bibitem[{\citenamefont{Hache} \emph{et~al.}(1986)\citenamefont{Hache, Ricard,
  and Flytzanis}}]{hache_86_1}
\bibinfo{author}{\bibfnamefont{F.}~\bibnamefont{Hache}},
  \bibinfo{author}{\bibfnamefont{D.}~\bibnamefont{Ricard}}, \bibnamefont{and}
  \bibinfo{author}{\bibfnamefont{C.}~\bibnamefont{Flytzanis}},
  \bibinfo{journal}{J. Opt. Soc. Am. B}
  \textbf{\bibinfo{volume}{3}}(\bibinfo{number}{12}), \bibinfo{pages}{1647}
  (\bibinfo{year}{1986}).

\bibitem[{\citenamefont{Rautian}(1997)}]{rautian_97_1}
\bibinfo{author}{\bibfnamefont{S.~G.} \bibnamefont{Rautian}},
  \bibinfo{journal}{J. Exp. Theor. Phys.}
  \textbf{\bibinfo{volume}{85}}(\bibinfo{number}{3}), \bibinfo{pages}{451}
  (\bibinfo{year}{1997}).

\bibitem[{\citenamefont{Pustovit and
  Shahbazyan}(2006{\natexlab{a}})}]{pustovit_06_1}
\bibinfo{author}{\bibfnamefont{V.~N.} \bibnamefont{Pustovit}} \bibnamefont{and}
  \bibinfo{author}{\bibfnamefont{T.~V.} \bibnamefont{Shahbazyan}},
  \bibinfo{journal}{J. Opt. Soc. Am. A}
  \textbf{\bibinfo{volume}{23}}(\bibinfo{number}{6}), \bibinfo{pages}{1369}
  (\bibinfo{year}{2006}{\natexlab{a}}).

\bibitem[{\citenamefont{Pustovit and
  Shahbazyan}(2006{\natexlab{b}})}]{pustovit_06_2}
\bibinfo{author}{\bibfnamefont{V.~N.} \bibnamefont{Pustovit}} \bibnamefont{and}
  \bibinfo{author}{\bibfnamefont{T.~V.} \bibnamefont{Shahbazyan}},
  \bibinfo{journal}{Phys. Rev. B} \textbf{\bibinfo{volume}{73}},
  \bibinfo{pages}{085408} (\bibinfo{year}{2006}{\natexlab{b}}).

\bibitem[{\citenamefont{Drachev} \emph{et~al.}(2004)\citenamefont{Drachev,
  Buin, Nakotte, and Shalaev}}]{drachev_04_2}
\bibinfo{author}{\bibfnamefont{V.~P.} \bibnamefont{Drachev}},
  \bibinfo{author}{\bibfnamefont{A.~K.} \bibnamefont{Buin}},
  \bibinfo{author}{\bibfnamefont{H.}~\bibnamefont{Nakotte}}, \bibnamefont{and}
  \bibinfo{author}{\bibfnamefont{V.~M.} \bibnamefont{Shalaev}},
  \bibinfo{journal}{Nano Letters}
  \textbf{\bibinfo{volume}{4}}(\bibinfo{number}{8}), \bibinfo{pages}{1535}
  (\bibinfo{year}{2004}).

\bibitem[{\citenamefont{Torres-Torres}
  \emph{et~al.}(2007)\citenamefont{Torres-Torres, Khomenko, Cheang-Wong,
  Rodriguez-Fernandez, Crespo-Sosa, and Oliver}}]{torres-torres_07_1}
\bibinfo{author}{\bibfnamefont{C.}~\bibnamefont{Torres-Torres}},
  \bibinfo{author}{\bibfnamefont{A.~V.} \bibnamefont{Khomenko}},
  \bibinfo{author}{\bibfnamefont{J.~C.} \bibnamefont{Cheang-Wong}},
  \bibinfo{author}{\bibfnamefont{L.}~\bibnamefont{Rodriguez-Fernandez}},
  \bibinfo{author}{\bibfnamefont{A.}~\bibnamefont{Crespo-Sosa}},
  \bibnamefont{and} \bibinfo{author}{\bibfnamefont{A.}~\bibnamefont{Oliver}},
  \bibinfo{journal}{Opt. Express}
  \textbf{\bibinfo{volume}{15}}(\bibinfo{number}{15}), \bibinfo{pages}{9248}
  (\bibinfo{year}{2007}).

\bibitem[{\citenamefont{Butenko} \emph{et~al.}(1990)\citenamefont{Butenko,
  Chubakov, Danilova, Karpov, Popov, Rautian, Safonov, Slabko, Shalaev, and
  Stockman}}]{butenko_90_1}
\bibinfo{author}{\bibfnamefont{A.~V.} \bibnamefont{Butenko}},
  \bibinfo{author}{\bibfnamefont{P.~A.} \bibnamefont{Chubakov}},
  \bibinfo{author}{\bibfnamefont{Y.~E.} \bibnamefont{Danilova}},
  \bibinfo{author}{\bibfnamefont{S.~V.} \bibnamefont{Karpov}},
  \bibinfo{author}{\bibfnamefont{A.~K.} \bibnamefont{Popov}},
  \bibinfo{author}{\bibfnamefont{S.~G.} \bibnamefont{Rautian}},
  \bibinfo{author}{\bibfnamefont{V.~P.} \bibnamefont{Safonov}},
  \bibinfo{author}{\bibfnamefont{V.~V.} \bibnamefont{Slabko}},
  \bibinfo{author}{\bibfnamefont{V.~M.} \bibnamefont{Shalaev}},
  \bibnamefont{and} \bibinfo{author}{\bibfnamefont{M.~I.}
  \bibnamefont{Stockman}}, \bibinfo{journal}{Z. Phys. D}
  \textbf{\bibinfo{volume}{17}}, \bibinfo{pages}{283} (\bibinfo{year}{1990}).

\bibitem[{\citenamefont{Danilova} \emph{et~al.}(1996)\citenamefont{Danilova,
  Drachev, Perminov, and Safonov}}]{danilova_96_2}
\bibinfo{author}{\bibfnamefont{Y.~E.} \bibnamefont{Danilova}},
  \bibinfo{author}{\bibfnamefont{V.~P.} \bibnamefont{Drachev}},
  \bibinfo{author}{\bibfnamefont{S.~V.} \bibnamefont{Perminov}},
  \bibnamefont{and} \bibinfo{author}{\bibfnamefont{V.~P.}
  \bibnamefont{Safonov}}, \bibinfo{journal}{Bulletin of the Russian Acad. Sci.
  - Physics} \textbf{\bibinfo{volume}{60}}(\bibinfo{number}{3}),
  \bibinfo{pages}{342} (\bibinfo{year}{1996}).

\bibitem[{\citenamefont{Danilova} \emph{et~al.}(1997)\citenamefont{Danilova,
  Lepeshkin, Rautian, and Safonov}}]{danilova_97_1}
\bibinfo{author}{\bibfnamefont{Y.~E.} \bibnamefont{Danilova}},
  \bibinfo{author}{\bibfnamefont{N.~N.} \bibnamefont{Lepeshkin}},
  \bibinfo{author}{\bibfnamefont{S.~G.} \bibnamefont{Rautian}},
  \bibnamefont{and} \bibinfo{author}{\bibfnamefont{V.~P.}
  \bibnamefont{Safonov}}, \bibinfo{journal}{Physica A}
  \textbf{\bibinfo{volume}{241}}, \bibinfo{pages}{231} (\bibinfo{year}{1997}).

\bibitem[{\citenamefont{Drachev} \emph{et~al.}(1998)\citenamefont{Drachev,
  Perminov, Rautian, and Safonov}}]{drachev_98_1}
\bibinfo{author}{\bibfnamefont{V.~P.} \bibnamefont{Drachev}},
  \bibinfo{author}{\bibfnamefont{S.~V.} \bibnamefont{Perminov}},
  \bibinfo{author}{\bibfnamefont{S.~G.} \bibnamefont{Rautian}},
  \bibnamefont{and} \bibinfo{author}{\bibfnamefont{V.~P.}
  \bibnamefont{Safonov}}, \bibinfo{journal}{JETP Lett.}
  \textbf{\bibinfo{volume}{68}}(\bibinfo{number}{8}), \bibinfo{pages}{651}
  (\bibinfo{year}{1998}).

\bibitem[{\citenamefont{Karpov} \emph{et~al.}(2001)\citenamefont{Karpov,
  Kodirov, Ryasiyanskiy, and Slabko}}]{karpov_01_1}
\bibinfo{author}{\bibfnamefont{S.~V.} \bibnamefont{Karpov}},
  \bibinfo{author}{\bibfnamefont{M.~K.} \bibnamefont{Kodirov}},
  \bibinfo{author}{\bibfnamefont{A.~I.} \bibnamefont{Ryasiyanskiy}},
  \bibnamefont{and} \bibinfo{author}{\bibfnamefont{V.~V.}
  \bibnamefont{Slabko}}, \bibinfo{journal}{Quantum Electronics}
  \textbf{\bibinfo{volume}{31}}(\bibinfo{number}{10}), \bibinfo{pages}{904}
  (\bibinfo{year}{2001}).

\bibitem[{\citenamefont{Boyd}(1992)}]{boyd_book_92}
\bibinfo{author}{\bibfnamefont{R.~W.} \bibnamefont{Boyd}},
  \emph{\bibinfo{title}{Nonlinear Optics}} (\bibinfo{publisher}{Academic
  Press}, \bibinfo{address}{Boston}, \bibinfo{year}{1992}).

\bibitem[{fn_()}]{fn_2}
 \bibinfo{note}{{More} general polarization can be considered
  separately.}

\end{thebibliography}

\end{document}